\documentclass[runningheads]{llncs}
\usepackage[T1]{fontenc}

\usepackage{graphicx}
\usepackage{mathtools}
\usepackage{amsfonts}
\usepackage{newtxtext,newtxmath}

\DeclareMathAlphabet{\mathbb}{U}{msb}{m}{n}
\begin{document}
\title{Semiflows, Home Spaces, and Home States,\\ Applications to the Analysis of Parameterized Petri nets}
%
%
\author{Gerard Memmi
}
\authorrunning{G. Memmi}
%
\institute{LTCI, Telecom-Paris, Institut polytechnique de Paris, Palaiseau, France
}
\maketitle              
\begin{abstract}
 
After rapidly recalling basic notations relatively to semiflows and Petri nets, we define $\mathcal{F}$, the set of semiflows over $\mathbb{Z}$ that we associate with a specific class of invariants. We then focus on $\mathcal{F^{+}}$, the set of semiflows with non-negative coordinates which are important to study the behavior of a Petri net. 
We recall known behavioral properties attached to semiflows in $\mathcal{F^{+}}$ that we associate with two sets of bounds regarding boundedness then liveness.
We recall the notions of home states and home spaces for which we regrouped old and new properties.
We introduce a new result on the decidability of liveness under the existence of a home state.

The notions of minimality of semiflows and minimality of supports allow us to define generating sets that are particularly critical to develop an effective analysis of invariants and behavioral properties of Petri nets such as boundedness or even liveness. 

We also recall three known decomposition theorems considering  
$\mathbb{N}$, 
$\mathbb{Q^+}$, and 
$\mathbb{Q}$ respectively 
where the decomposition over $\mathbb{N}$ is being improved with a necessary and sufficient condition.

We use the notion the notion of generating sets to show that extremums linked to the set of bounds mentioned here above, are indeed computable by providing their values.

Two related Petri net modeling arithmetic operations (one of which represents an Euclidean division) illustrate how results on semiflows and home spaces can be methodically used in efficiently analyzing the liveness of the parameterized model and underlining the efficiency brought by the combination of these results. 
 
\end{abstract}

\section{Introduction}

\subsection{Motivations}
Parallel programs, distributed digital systems, telecommunication networks, or cyber-physical systems are  entities that are complex to design, model, and verify. 
Using formal verification at different stages of the system development life cycle is a strong motivation and provides us with the rationale for revisiting the notions of semiflows and home spaces and illustrate through examples how they can be combined to analyze behavioral properties of Petri nets.
In this regard, invariants are of paramount importance as they are almost systematically used in system specifications to describe specific behavioral properties. One can argue that properties such as liveness, deadlock freeness, or boundedness are in some way invariants since they must hold regardless of the evolution of the digital system under study.


Often, engineers and researchers will try to prove that a formula belonging to a system specification is an invariant, meaning that the formula holds during any possible evolution of their model. 
But, can we find a way by which invariants or at least a meaningful subset of invariants can be organized and concisely described, and some of them can be discovered by computation? Such invariants that do not belong in the system specification, can just express a sub-property of a more complex known one; however, they also can reveal an under-specified model or an unsuspected function of the system under study (which in turn, could constitute a component of a security breach).
In this paper, we provide some elements to answer this question and show how basic arithmetic, linear algebra, or algebraic geometry can efficiently support invariant calculus.

One of our motivations is to go beyond regrouping in a kind of survey, a number of known algebraic results dispersed throughout the Petri nets literature, but also to introduce improved and new results.
We illustrate how engineers can methodically proceed in their analysis through two parameterized Petri nets.

This paper can be considered as a continuation of the work started in \cite{M23} and an extensive rewriting of \cite{Me25}, providing in particular a better positioning between semiflows and and its associated set of invariants through theorem (\ref{th: inv-flow}). A conjecture proposed in \cite{Me25} is solved for state machines, a subclass of Petri nets although negatively negatively for the general case through a counterexample.
We want to show how linear algebra or algebraic geometry can efficiently be applied and utilized to prove a large variety of behavioral properties, sometimes with simple arithmetic reasoning even when Petri net or colored Petri nets (\cite{JR97}) are parameterized.
This type of methodical reasoning can be useful to engineers in order to determine in which domain of these parameters, behavioral properties can be satisfied.




\subsection{Outline and contributions}
After providing basic notations in Section \ref{sec: basic notations}, we regroup a first set of classic properties for semiflows (over $\mathbb{Z}$ then $\mathbb{N}$) and introduce the definitions of its associated family of invariants (theorem \ref{th: inv-flow}) as well as two first extremums in Section \ref{sec: semi}.

The notions of home space and home state are described in Section \ref{sec: HS} together with a set of old and new results linked to the topology of the underlying reachability graph.
Later, in Section \ref{subsec: hs-liveness}, their key relation with liveness is described highlighting their usefulness. 
In particular, a new decidability result is provided for Petri nets with home states linked to Karp and Miller's coverability tree finite construction \cite{KM69} through theorem \ref{th: home-state} and corollary \ref{cor: home-state}.
In Section \ref{subsec: semiflows and home spaces} a third extremum is defined. 

The notions of generating sets and minimality are briefly recalled from \cite{M23}
%
in Section \ref{sec: generating sets}. 
The three decomposition theorems of Section \ref{subsec: 3theorems} have been first published in \cite{M78} then improved in \cite{M23}. 
Here, the first theorem is extended once more to fully characterize minimal semiflows and generating sets over $\mathbb{N}$. The other two theorems are just recalled for completeness.
 


Subsequently, Section \ref{subsec: bounds}, illustrating  the usefulness of decomposition theorems, theorem \ref{th: bounds} shows how to compute the three extremums previously defined when equipped with a generating set. 
It is worth noticing that the results do not depend on the chosen generating set. 
These important details were never stressed out before despite their importance from a computational point of view, and their impact in supporting the analysis of parameterized models. 

These results are used in the analysis of two examples presented in Section~\ref{sec: ex} where two parameterized examples are given to illustrate how invariants and home spaces can be associated together with basic arithmetic reasoning to methodically prove behavioral properties of a Petri net (described section \ref{subsec: method}).
This section ends with a last result on home states directly drawn from the analysis of the last example.

Section \ref{sec: concl} concludes and provides a possible avenue for future research.

\section{Petri nets, basic notations}
\label{sec: basic notations}
In this section, after describing basic mathematical notations that will be used in this paper, we briefly recall Petri nets basic definitions, including the notion of potential state space that is more usual in Transition Systems. 

\subsection{Notations}
$\mathbb{S}$ denotes an element of $ \{\mathbb{N}, \mathbb{Q^{+}}, \mathbb{Q}\}$ where $\mathbb{Q^{+}}$ is denoting the set of non-negative rational numbers.

Given an ordered set $A = \{a1, ...an\}$, a vector $v = (x1,...xn)^\top$ with its coordinates in $\mathbb{S}$ can also be seen as a function from $A$ to $\mathbb{S}$ with $v(ai)= xi$ sometimes also denoted by $v_i$.

The \textit{support} of a vector $v = (x1,...xn)^\top$ is denoted by $\left \| v \right \|$ and is defined by:
$$\left \| v \right \| = \{ai \in A |\ v(ai) \neq 0\}.$$

We will abusively use the same symbol ‘0' to denote 
$(0,...,0)^\top$ of $\mathbb{N}{^n}$,\ for all $n$ in $\mathbb{N}$. 
We will use the usual component-wise 
partial orders ($\leq \text{or} <$) between vectors in which $(x_1, x_2, \dots, x_d )^\top \leq ( y_1, y_2, \dots, y_d )^\top$ if and only if $x_i \leq y_i$, for all $i \in \{1, \dots, d\}$.

These notations and conventions will be useful to handle the notion of semiflow in section \ref{sec: semi}.
\subsection{Petri nets}
\label{subsec: Petri nets definitions}

A \textit{Petri net} is a tuple $PN = \langle P, T, Pre, Post \rangle$, where $P$ is a finite set of \textit{places} and $T$ a finite set of \textit{transitions} such that $P \cap T = \text{\O}$. 
A transition $t$ of $T$ is defined by its $\mathit{Pre(\cdot,t)}$ and $Post(\cdot,t)$ \textit{conditions}\footnote{We use here the usual notation: $Pre(\cdot,t)(p) = Pre(p,t)$ and  $Post(\cdot,t)(p) = Post(p,t)$.}:
$Pre: P \times T \rightarrow \mathbb{N}$ is a function providing a weight for pairs ordered from places to transitions, while $Post: P \times T \rightarrow \mathbb{N}$ is a function providing a weight for pairs ordered from transitions to places. 
Here, $d$ will denote the number of places: $d = |P|$.
A Petri net is said \textit{ordinary} \cite{STC98} when $Pre$ and $Post$ are functions from $P \times T$ to $\{0,1\}$.

The \textit{structure} of a Petri net as a bipartite graph, is often studied in using the notion of siphons and traps (see for instance, \cite{Colom2003}) which are useful to analyze behavioral properties.
A \textit{siphon} $D$ is a subset of places such that $\forall t \in T,\ \forall p \in D,$ if $Post(p,t) \neq 0$ then $\exists p' \in D$ such that $Pre(p',t) \neq 0$. Similarly, a \textit{trap} D is a subset of places such that $\forall t \in T,\ \forall p \in D,$ if $Pre(p,t) \neq 0$ then $\exists p' \in D$ such that $Post(p',t) \neq 0$. 
The structure of a Petri net can also be constrained to characterize subclasses of Petri nets. As an example, States Machines are a particularly well-known subclass of Petri nets usually used to model sequential processes. A \textit{state machine} $PN = \langle P, T, Pre, Post \rangle$ is an ordinary Petri net such that: 
$$\forall t \in T, \ \sum_{p \in P} Pre(p,t) = \sum_{p \in P} Post(p,t) = 1$$

A \textit{marking} (or \textit{state} in Transition Systems) $q:\ P \to \mathbb{N}$ allows representing the evolution of the system along the execution (or \textit{firing}) of a transition $t$ or of a sequence of transitions $\sigma$ (i.e., a word in $T^*$).
Given a marking $q$, a place $p$ is said to contain $k$ \textit{tokens} as $q(p) = k$.

We say that $t$ is \textit{enabled at} marking $q$ if and only if $q \geq Pre(\cdot, t),$
and only an enabled transition $t$ at $q$ (we write that $q \in \texttt{Dom}(t)$) can be executed, reaching a marking $q'$ computable from $q$
such that:
$$q' = q + Post(\cdot, t) - Pre(\cdot, t)$$

This is also denoted as $q' =t(q)$ or more traditionally 
$q \overset{t}{\rightarrow}q'$ (we also write $q' \in \texttt{Im}(t)$).
Similarly, for a sequence of transitions $\sigma$ allowing to reach a marking $q'$ from a marking $q$, we write $q \overset{\sigma}{\rightarrow}q'$.
If $\sigma$ can be written $\sigma = t_1t_2...t_n$ then we can compute $q'= t_n(...t_1(q)...)$ as the composition of the successive transitions of $\sigma$ with the so-called \textit{state equation}:
\begin{equation}
\label{eq: state}
    q' = q + \sum_{i=1}^{i=n}\Bar{\sigma_i}(Post(\cdot, t_i) - Pre(\cdot, t_i)) = q + (Post-Pre)\Bar{\sigma}
\end{equation}
where Post and Pre are to be considered as matrices (often, Post-Pre is defined as the \textit{incidence matrix} of $PN$), $\Bar{\sigma}_i$ is the number of occurrences of $t_i$ in $\sigma$, and $\Bar{\sigma}$ is the Parikh vector of $\sigma$ that is the vector of $\mathbb{N}^{|T|}$ such that $\Bar{\sigma}_i$ is the $i^{th}$ coordinate of $\sigma$.

When the sequence of transitions allowing to reach a marking $q'$ from a marking $q$ is unknown, we may write $q \overset{*}{\rightarrow}q'$.
In some papers, $\overset{*}{\rightarrow}$ denotes the reachability relation between markings.

We also define $Q$, the set of all \textit{potential markings} also known as \textit{potential state space} or state space in Transition Systems. 
Without additional information on the domain in which marking of places may vary, we will assume 
$Q = \mathbb{N}^d$.
Sometimes, information about the system under study allows to reduce $Q$; in that case, we will assume to have at least $\forall t \in T, Pre(\cdot,t)\in Q\ \text{and}\ Post (\cdot, t) \in Q$ hold. 

$RS(PN,Init)$ denotes the set of reachability of a Petri net $PN$ from a subset $Init$ of $Q$: 
$RS(PN,Init) = \{q \in Q \ |\ \exists\ a \in Init,\ a\overset{*}{\rightarrow} q \}$.

$RG(PN,Init)$, and $LRG(PN,Init)$ denote the reachability graph without labels (as in Figure \ref{fig: inter-hs}) and with labels in $T$ respectively; while $LCT(PN,q_0)$ denote the labeled coverability tree given a single initial marking $q_0$.


\section{Semiflow and Invariants, basic properties}
\label{sec: semi}

The concept of semiflows over non negative integers were first described by Y.E. Lien \cite{L73} and independently by K. Lautenbach and H. A. Schmid \cite{LS74}. 
The algebraic calculus underneath can be find in \cite{M77}. Then, M. Silva \cite{martinez1982} extended the definition to semiflows over integers.
After recalling the definition of semiflows, we gather four properties illustrating their their link with behavioral properties.

In this section, we consider $\langle PN,q_0 \rangle$, a Petri net $PN$ with its initial marking $q_0$ and the set of reachable markings from $q_0$ through all sequences of transitions denoted by $RS(PN,q_0)$.
We first give a classic definition of a semiflow, then show how they characterize a class of invariants that we call \textit{f-q-invariant}.
Then, we analyze how semiflows relate to boundedness as well as liveness.

\subsection{Semiflows and associated invariants}
\label{subsec: semi-inv}
\begin{definition} [Semiflow]
A \textit{Semiflow} $f$ is a solution of the following homogeneous system of $|T|$ diophantine equations: 
\begin{equation}
\label{eq: inv-semiflow}
    f^\top Post(\cdot,t) = f^\top Pre(\cdot,t), \ \ \forall t \in T,
\end{equation}
where $x^\top y$ denotes the scalar product of the two vectors $x$ and $y$, since $f, Pre(\cdot,t)$ and $Post(\cdot,t)$ can be considered as vectors once the places of $P$ have been ordered. 

$\mathcal{F}$ and $\mathcal{F}^+$ denote the sets of solutions of the system of equations (\ref{eq: inv-semiflow}) that have their coefficients in $\mathbb{Z}$ and in $\mathbb{N}$, respectively.
\end{definition}


The following known property (\cite{M78}, \cite{BR82}, or \cite{STC98}) can easily be proven true in $\mathbb{N}^d$ and not true in $\mathcal{F}$: 
\begin{property}
\label{prop: support-plus-union}
If $f$ and $g$ are two vectors with non-negative coefficients, then we have:
$\left\|f+g \right\|=\left\|f \right\|\cup \left\| g\right\|$.

If $\alpha$ is a non-null integer then $\left\|\alpha f \right\|=\left\|f \right\|$.

If $f \in \mathcal{F}^+$, then $\left\|f \right\|$ is both a siphon and a trap which is not the case when $f \in \mathcal{F} \setminus \mathcal{F}^+$.
\end{property}
Even when focusing on Petri net literature, we can find many different definitions of the notion of invariant. 
For instance, an invariant is a set of markings in \cite{Si82,L09,JALE22}. 
Often, this notion is not even informally defined.
In this paper, an \textit{invariant} $I$ is a logic formula that is true for all markings $q'$ reached from $\langle PN,q \rangle$.
We write $\forall q' \in RS(PN,q), q \models I$
and say that $q$ \textit{satisfies} $I$.

Here, we are interested in a specific class of invariant that are directly attached to the notion of semiflow.
\begin{definition}[f-q-invariant]
\label{def: inv}
    Given $f \in \mathbb{Z}^d, I_{f,q} : Q \to \mathbb{F}_2$ denotes the following unary logic formula: $I_{f,q}(q'): f^\top q = f^\top q'$

    $I_{f,q}$ is an f-q-invariant if and only if
    \begin{equation}
\label{eq: inv}
\forall \ q'\ \in RS(PN,q): 
    f^\top q = f^\top q'.
\end{equation}
The \textit{support} of a logic formula $I_{f,q}$ sometimes also called the characteristic set of $I_{f,q}$ is the subset of $Q$ for which $I_{f,q}$ is true\footnote{In a more general manner, we sometimes say that the support $H(f,q)$ of $I_{f,q}$ is the subset of $Q$ that satisfies $I_{f,q}$ and write $H(f,q) = \{q' \in Q \ |\  q' \models I_{f,q} \}$.} and is denoted by: $$H(f,q) = \{q' \in Q \ |\ f^\top q = f^\top q'\}.$$
\end{definition}
Indeed, we have $\forall q \in Q, (H(0,q)= Q$ and $\forall f \in \mathbb{Z}^d, q \in H(f,q)$).

The first argument of the support $H(f,q)$ of an f-q-invariant does not have a linear property; instead, we have: 
\begin{property}
\label{prop: inter-home-spaces}
    If $f \in \mathbb{Z}^d$, then,
    for all $ \alpha \in \mathbb{Q} \setminus \{0\}, H(\alpha f,q_0)= H(f,q_0)$. Also, for all $ f$ and $g \in \mathcal{F}$ and for all $ \alpha$ and $\beta \in \mathbb{Q}, H(f,q_0) \cap H(g, q_0) \subseteq H(\alpha f+ \beta g,q_0)$. 
\end{property}
If $q \in H(f,q_0) \cap H(g,q_0)$, then $\alpha (f^\top q) = \alpha (f^\top q_0)$ and $\beta (g^\top q) = \beta (g^\top q_0)$, 
so $(\alpha f+\beta g)^\top q = (\alpha f+ \beta g)^\top q_0$, and, therefore, $q \in H(\alpha f+\beta g,q_0)$
\hfill 
$\square$

Of course, $I_{f,q}$ is an f-q-invariant if and only if $RS(PN,q) \subseteq H(f,q)$. f-q invariants and semiflows are indeed related to each other thanks to the fact that a solution of equation (\ref{eq: inv-semiflow}) necessarily  satisfies equation \ref{eq: inv}. 
Then, the two following results make this interrelation more precise.
\begin{lemma}
\label{lem: semi-inv}
    Given a marking $qi$, if $f$ is a semiflow then $I_{f,qi}$ is an f-q-invariant.
\end{lemma}
If $f$ is a semiflow, it satisfies the system of equations (\ref{eq: inv-semiflow}), therefore, we can easily deduce from equation (\ref{eq: state}) that equations (\ref{eq: inv})
\hfill
$\square$

\begin{figure}[ht]
\centering
\includegraphics[width=0.36\textwidth]{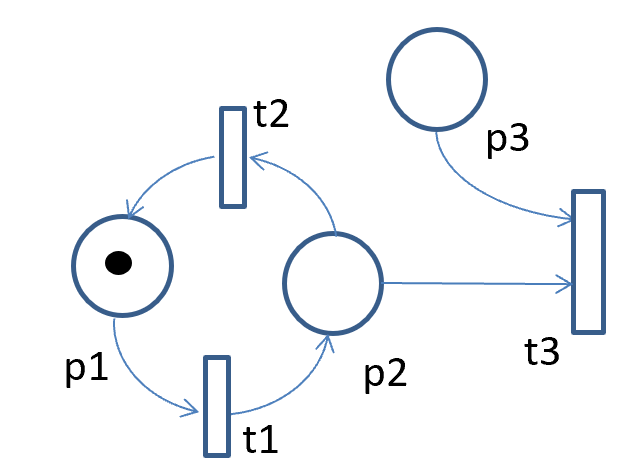}
\caption{
Considering the initial marking $qi^\top = (1,0,0)$, and  $f^\top = (1,1,0)$, we can easily verify that $I_{f,qi}$ is an f-q-invariant since $RS(PN,qi) = \{qi, (0,1,0)^\top\}$.\\ 
However, $f$ is not a semiflow, since it does satisfy the equation associated to $t1$ of the system of equations (\ref{eq: inv-semiflow}) since we have: $f^\top Pre(\cdot,t1)=1$ and $f^\top Post(\cdot,t1) = 0$.}
\label{fig: f-q-inv}
\end{figure}

Let us point out that the reverse is not true: if $I_{f,qi}$ is an f-q-invariant then f is not necessarily a semiflow as easily shown Figure (\ref{fig: f-q-inv}).

Actually, a semiflow can be characterized by the following theorem:
\begin{theorem}
\label{th: inv-flow}
$f$ is a semiflow if and only if $\forall q \in Q, I_{f,q}$ is an f-q-invariant
\end{theorem}
Given transition $t$ by choosing $Pre(\cdot,t)$ as initial marking \footnote{we assume section \ref{sec: basic notations} that $\forall t \in T, Pre(\cdot,t) \in Q$.}, we can execute $t$ and reach the marking $Post(\cdot,t)$. 
If $\forall q \in Q, I_{f,q}$ is an f-q-invariant then by equation (\ref{eq: inv}), we have: $f^\top Post(\cdot,t) = f^\top Pre(\cdot, t)$. We can iterate this operation for all transition of $T$, therefore, $f$ satisfies the system (\ref{eq: inv-semiflow}) and is a semiflow.
The reverse is obtained by applying lemma \ref{lem: semi-inv}
\hfill
$\square$

\subsection{Semiflows, boundedness, and liveness}
\label{subsec: semi-bound-live}
The most interesting set of semiflows, from a behavioral analysis standpoint, is $\mathcal{F}^+$, defined over natural numbers. This can be seen through the following three properties. 
First, we define the \textit{positive and negative supports} of a semiflow $f \in \mathcal{F}$ as:
$$\left\| f\right\|_+ = \left\{ p\in P\ |\ f(p)> 0\right\}$$
and $$
\left\| f\right\|_- = \left\{ p\in P\ |\ f(p)< 0\right\},$$
with $\left\| f\right\| = \left\| f\right\|_- \cup\   \left\| f\right\|_+$ (by property \ref{prop: support-plus-union}).
We can then rewrite Equation (\ref{eq: inv}) as:
\begin{equation}
\label{eq: differential}
\forall q \in RS(PN,q_0):\ \ f^\top q=\left| \sum_{p \in \left\| f\right\|_+}f(p)q(p)\right| - \left| \sum_{p \in \left\| f\right\|_-}f(p)q(p)\right|=f^\top q_0.
\end{equation}

As we can see, the formulation of Equation (\ref{eq: differential}) is a subtraction between the weighted number of tokens in the places belonging to the positive support and the weighted number of tokens in the places belonging to the negative support of $f$.

Let us recall that a subset of places $A$ is said ``bounded" in $\langle PN,q_0 \rangle$ if and only if 
$\forall p \in A,\ \exists k \in \mathbb{N}$ such that $\forall q \in RS(PN,q_0),\ q(p) \leq k$.
A first general property can be immediately deduced by recalling that any marking $q$ belongs to $\mathbb{N}^d$.

\begin{property}
\label{prop: boundedness in F} 
    For any semiflow $f \in \mathcal{F}$,
    $\left\| f\right\|_+$ is bounded if and only if $\ \left\| f\right\|_-$ is bounded.
\end{property}

Furthermore, equation \eqref{eq: differential} provides a convenient framework for characterizing the concept of ``implicit place" as introduced by Silva in \cite{silva80} for ordinary Petri nets (see definition section \ref{subsec: Petri nets definitions}). 
\begin{definition}
   In an ordinary Petri net $\langle PN,q_0\rangle$ a place $pi$ is said to be \textit{implicit} if and only if there exists a semiflow $f \in \mathcal{F}$ such that $$\left\| f \right\|_- = \{pi\},\  f(pi) = -1,\  \left\| f \right\|_+ \neq \text{\O}\  \textit{and}\  f^\top q_0 = 0.$$ 
\end{definition}
It is important to point out that in order to satisfy equation \eqref{eq: inv-semiflow}, we necessary have: $\forall t \in T$ such that $Pre(pi, t) = 1, \exists p' \in \left\| f \right\|_+$ such that $Pre(p',t) = 1$.
If $\exists q \in RS(PN,q_0)$ such that $q(pi) = 0$ then by equation \eqref{eq: differential} and the fact that PN is ``ordinary" $\forall p \in \left\| f \right\|_+, q(p) = 0$, in particular, $q(p') = 0$.
This is making the place $pi$ redundant from a behavioral point of view and $pi$ can be removed from PN without changing $LRG(PN,q_0)$.

On another hand, let us notice that the marking of $pi$ is fully determined by the markings of the places of $\left\| f \right\|_+$.

A more general algebraic definition can be found in \cite{BR76} with more general $Pre$ and $Post$ functions.

A behavioral definition can be found in \cite{Haddad88,SCC92,GVC99} or in \cite{T24}, where a place $pi$ is implicit if and only if removing $pi$ from $PN$ does not change its associated labeled reachability graph; which is the actual motivation for defining the concept of implicit place.

Of course, if $\left\| f\right\|_- =\ $\O, then $f \in \mathcal{F}^+$ and $\left\| f\right\|$ is necessarily \textit{structurally bounded} (i.e., bounded from any initial marking) and is sometime called \textit{conservative component} as in \cite{L76}. 
More generally, considering a weighting function $f$ over $P$  being defined over non-negative integers and verifying the following system of inequalities: 
\begin{equation}
\label{eq: upper-bound}
    f^\top Post(\cdot,t) \leq f^\top Pre(\cdot,t), \ \ \forall t \in T
\end{equation}
 
\begin{property}
\label{prop: boundedness-alg}
If $f \geq 0$ is such that it verifies the system of inequalities (\ref{eq: upper-bound}),
then the set of places of $\left\|f \right\|$ is structurally bounded.
Moreover, if $f \neq 0$ then $\left\|f \right\|$ is a siphon. 

Furthermore, the marking of any given place $p$ of $\left \| f \right \|$ has an upper bound:
$$q(p) \leq \left \lfloor \frac{f^{T}q_0}{f(p)} \right \rfloor , \  \ \forall q \in RS(PN,q_0).$$
\end{property}
The first and last items of this property can be easily proven (see for example \cite{M78}).
If for a given place $p \in \left \| f \right \|$, we have a transition $t$ such that $Post(p,t) > 0$ then since $f \geq 0$ $f^\top Post(p,t) > 0$. 
By the system of inequalities (\ref{eq: upper-bound}), we have $f^\top Pre(p,t) > 0$; therefore, $\exists p' \in \left \| f \right \|,$ such that $pre(p',t) > 0$. 

Hence $\left \| f \right \|$ is a siphon as defined section \ref{subsec: Petri nets definitions}
\hfill
$\square$

The reverse is not true in general; it is easy to build a Petri net containing a siphon for which there is no non-null solution $f$ of the system of inequalities (\ref{eq: upper-bound}) such that $f \geq 0$.

If $f > 0$, then $\left\|f \right\|_+ = \left\|f \right\| = P$, and we say that the Petri net is also structurally bounded. 
The reverse is also true: if the Petri net is structurally bounded, then there exists a strictly positive solution for the system of inequalities above (see \cite{Si78} or \cite{BR82}). 
This property is obviously false for a semiflow that satisfies Equation (\ref{eq: inv-semiflow}) but would have at least one negative coordinate, and constitutes a first reason for particularly considering weight functions $f$ over $P$  being defined over non-negative integers including $\mathcal{F}^+$.
We can then define $\Lambda$ relatively to a given place $p$ and an initial marking $q_0$, the set of all possible bounds generated by semiflows in $\mathcal{F}^+$:
$$\Lambda(p,q_0) = \{x \in \mathbb{N} \ | \ \exists f \in \mathcal{F}^+, x= \left \lfloor \frac{f^{T}q_0}{f(p)} \right \rfloor\}$$ and its extremum (and more useful element): $$\lambda(p,q_0) = \min_{\{f\in \mathcal{F}^+\ |\ f(p) \neq 0\}} \left \lfloor \frac{f^\top q_0}{f(p)}\right \rfloor$$

The following corollary can be directly deduced from the fact that any semiflow in $\mathcal{F}^+$ satisfies the system of inequalities (\ref{eq: upper-bound}); therefore, Property \ref{prop: boundedness-alg} can apply:
\begin{corollary}
\label{cor: bound-for-support}
For any place $p$ belonging to at least one support of a semiflow of $\mathcal{F}^+$, an upper bound $\lambda$ can be defined for the marking of $p$ relatively to an initial marking $q_0$ such that:   
$$\forall q \in RS(PN,q_0), \ \ q(p) \leq \lambda(p,q_0) =  \min_{\{f\in \mathcal{F}^+\ |\ f(p) \neq 0\}} \left \lfloor\frac{f^\top q_0}{f(p)}\right \rfloor.$$
\end{corollary}

We will see with Theorem \ref{th: bounds} that this bound is computable as a consequence of Theorems \ref{th: over N} and \ref{th: decomp}.

\begin{definition}
\label{def: enabling-threshold}
    Given a transition $t$ and a semiflow $f$ in $\mathcal{F}^+$, the scalar product $f^{T} Pre(\cdot,t)$ is called the f-enabling threshold of $t$.
\end{definition}
Sometimes, when there is no ambiguity,  $f^{T} Pre(\cdot,t)$ is more simply called the \textit{enabling threshold} of $t$ as in \cite{BR82}. 

\begin{figure}[ht]
\centering
\includegraphics[width=0.3\textwidth]{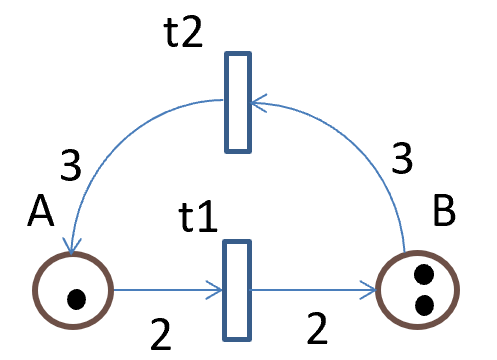}
\caption{$\mathcal{F}^+ = \{ (x,x)^\top | x \in \mathbb{N}\}$, $Pre(A,t1)=2, Pre(B,t1)=0, \ q_0(A)=1,\ q_0(B)=2$. \\
We can verify  that $1 \leq \theta(t1,q_0) = \min_{f \in \mathcal{F}^+} \frac{f^\top q_0}{f^\top Pre(\cdot,t1)}=3/2$. \\Yet, $t1$ cannot be executed from $q_0$.}
\label{fig: CS-threshold}
\end{figure}

This gives us a second reason for particularly considering a semiflow $f$
as being defined over non-negative integers is that the system of inequalities
\begin{equation}
\label{eq: enabling-threshold}
 f^{T} q_0  \geq f^{T} Pre(\cdot,t), \ \ \ \ 
 \forall t \in T,
\end{equation}
becomes a necessary condition for any transition $t$ to stand a chance to be enabled from a reachable marking from $q_0$, therefore to be live. 
Indeed, the system of inequalities (\ref{eq: enabling-threshold}) motivates the definition \ref{def: enabling-threshold}.
\begin{property}
\label{prop: enabling-threshold}
If $t$ is a transition and $f \in \mathcal{F}^+ \setminus \{0\}$ such that
 $f^{T} q_0  < f^{T} Pre(\cdot,t)$, then $t$ cannot be executed from $\langle PN,q_0 \rangle$.

 More generally, a necessary condition for a transition $t$ to be executed at least once from $\langle PN,q_0 \rangle$ is
 $$1 \leq \min_{\{f\in \mathcal{F}^+\ |\ f^\top Pre(\cdot, t) \neq 0\}} \frac{f^\top q_0}{f^\top Pre(\cdot, t)}$$
\end{property}
We can define $\Theta$ stemming from property \ref{prop: enabling-threshold} the same way we defined $\Lambda$ stemming from property \ref{prop: boundedness-alg}:
$$\Theta(t,q_0) = \{x \in \mathbb{Q}^+ \ |\ \exists f \in \mathcal{F}^+, \ x= \frac{f^\top q_0}{f^\top Pre(\cdot, t)}\}$$
with its extremum $\theta$ the inequality of property \ref{prop: enabling-threshold} can be rewritten such that:
$$1 \leq \theta(t,q_0) = \min_{\{x \in \Theta\}} x$$
Property \ref{prop: enabling-threshold} is not a sufficient condition for a transition to be enabled, see Figure \ref{fig: CS-threshold}.

Property \ref{prop: enabling-threshold} is of interest when the model is defined with parameters, since some values of these parameters for which the model is not live can be rapidly pruned away (see example Figure \ref{fig: tinyi}).
Moreover, it also says that the only way (without changing the structure of the model) to make the execution of $t$
possible is by adding tokens in $\left\| f  \right\|$ for any semiflow $f$ for which inequality (\ref{eq: enabling-threshold}) is not satisfied.
We conjecture that the notion of siphon \cite{Colom2003} could be useful here to infer a better bound for the marking of a place than the enabling threshold of property \ref{def: enabling-threshold}.

This property is used to prove theorem \ref{th: decomp} section \ref{subsec: 3theorems} and theorem \ref{th: bounds} section \ref{subsec: bounds}.
These results have been cited and utilized many times in various applications going beyond computer science, electrical engineering, or software engineering. 
For instance, they have been used in domains such as population protocols \cite{CzernerEL23} or biomolecular chemistry relative to chemical reaction networks \cite{JACB18}, which brings us back to the C. A. Petri's original vision, when he highlighted that his nets could be used in chemistry. Many other applications can be found in the literature.

\section{Home spaces and home states}
\label{sec: HS}

The notion of home space was first defined in \cite{M83} for Petri nets relatively to a single initial marking. 
Early descriptions of theorem \ref{th: johnen}, corollary \ref{cor: sink-home-space}, and properties \ref{prop: home-state}, \ref{prop: liveness-domain-t}, and \ref{prop: home-state-liveness} can be found in \cite{M83,J87,JM91} for a single initial marking.
Here, we effortlessly extended its definition relatively to a nonempty subset $Init$ of initial markings.
Furthermore, these results are also generalized by extension or addition of sub-properties.

Home spaces are extremely useful to analyze liveness (see \cite{VautherinM84}) or resilience (see \cite{FinHil24}). 
Any behavioral property requiring to potentially become satisfied after executing a sequence of transitions can be supported by a home space (i.e. have a home space for support). 
A property satisfied for any reachable marking would be an invariant.
A property satisfied for any marking of $Q$ would be a kind of tautology.

\subsection{Definitions and basic properties}
\label{subsec: HS-def}
Given a Petri net $PN$, its associated set $Q$ of all potential markings and a subset $Init$ of $Q$, we say that a set \emph{H} is an \textit {Init-home space} if and only if, for any progression (i.e. sequence of transitions) from any element of $Init$, there exists a way of prolonging this progression and reach an element of \emph{H}. In other words:

\begin{definition} [Home space]
\label{def: homespace}
Given a nonempty subset $Init$ of $Q$, a set $\emph{H}$ is an \textit{Init-home space} if and only if, for all $ q\in RS(PN,Init), RS(PN,q) \cap H \neq \text{\O}$, in other words, there exists $ h \in \emph{H}$ such that $h$ is reachable from $q$, (i.e. $q \overset{*}{\rightarrow}h$).

\end{definition} 
This definition is general and can be applied to any Transition System. In \cite{JALE22}, we can find, for Petri nets, an equivalent definition: $\emph{H}$ is an \textit{Init-home space} if and only if $RS(PN,Init) \subseteq RS^{-1}(PN,H \cap Q)$.

 

\begin{definition} [Home state]
\label{def: home state}
Given a nonempty subset \textit{Init} of $Q$, a marking $s$ is an \textit{Init-home state} if and only if 
$\{s\}$ is an \textit{Init-home space}. 
\end{definition}
If $s$ is an \textit{Init}-home state, then it is an $\{s\}$-home state, since  $RS(PN,s) \subseteq RS(PN,Init) \subseteq RS_1(PN,s)$.
We simply say that $s$ is a home state when there is no ambiguity. This is the usual notation that can be found in \cite{BR82}, p.59, in \cite{GV03}, p. 63, 
or in \cite{HDMK14} and in many other papers.

In many systems, there exists a unique \textit{idle} state from which the various capabilities of the system can be executed. 
Then, this initial state can be modeled by an initial marking $q_0$. 
In this case, it is important for $q_0$ to be a home state (i.e. for the system to always be able to go back to its initial state).
This property is usually guaranteed by a \textit{reset function} that can be modeled in a simplistic way by adding a transition $r$ such that $RS(PN,q_0) \subseteq \texttt{Dom}(r)$ and $\{q_0\} = \texttt{Im}(r)$ (which means that $r$ is executable from any reachable marking and that its execution reaches $q_0$, forcing $q_0$ to be a home state). 
However, most of the time this function is abstracted away from the model since it is adding too much complexity to $RG$ (one edge per node).

It is not always easy to prove that a given set $H$ is an Init-home space. 
This question is proven decidable in \cite{JALE22} and to be Ackermann-complete later in \cite{JanLe24} for Petri nets but is still open in for more complex conceptual model such as Transition Systems. Furthermore, a corpus of decidable properties regarding home spaces as well as home states, can be found in \cite{J87,EJ89,BESP16,FinHil24}, or \cite{JanLe24}. 

\begin{figure}[ht]
\centering
\includegraphics[width=0.54\textwidth]{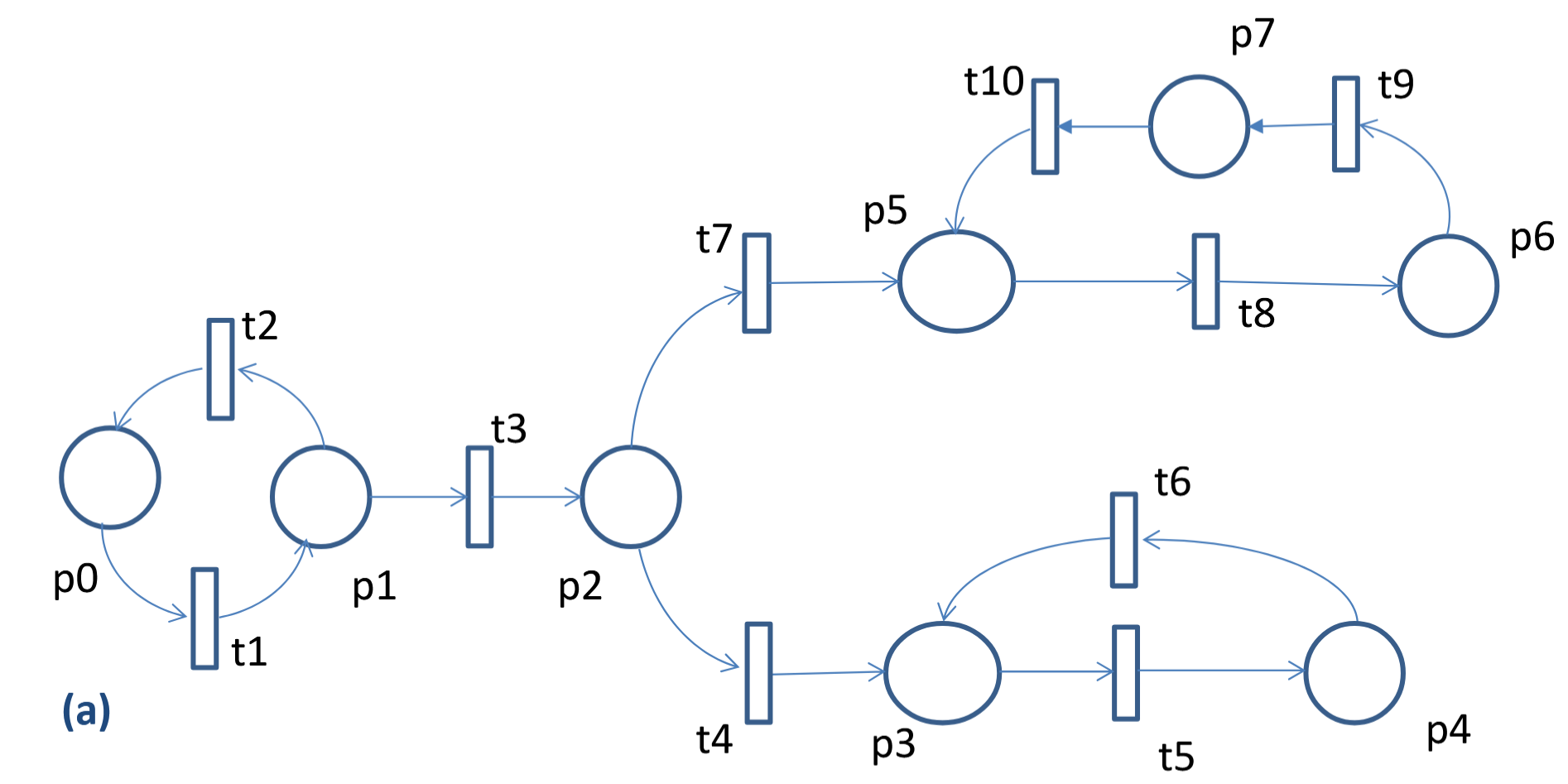}
\includegraphics[width=0.45\textwidth]{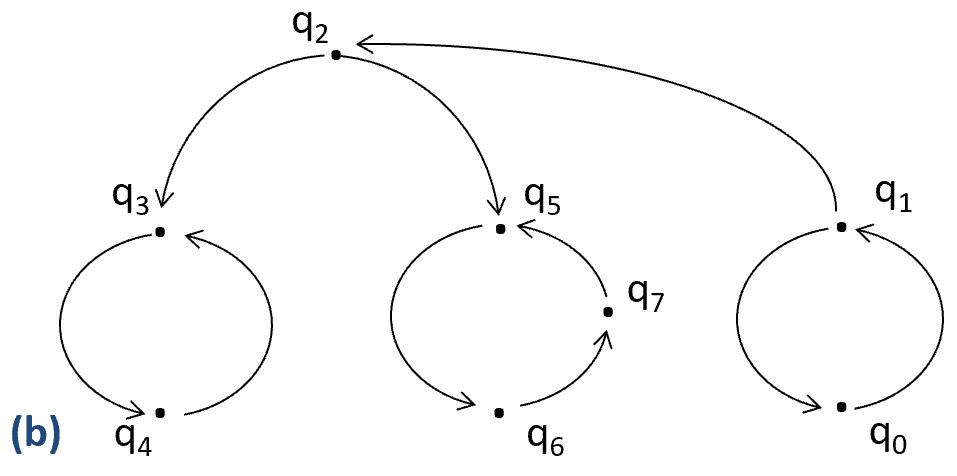}
\caption{
The Petri net $PN$ represented in \textbf{(a)} is a state machine such that $Init = \{q_0,q_1\}$,
and its Reachability Graph $RG(PN,Init)$ is represented in \textbf{(b)} where $q_i(pi) = 1, q_i(pj) = 0,\ \text{for}\ i \neq j, \forall i,j \in \{0,...7\}$. \\ 
$H_1=\{q_3,q_5,q_6\}$, $H_2=\{q_3,q_7\}$, and $H_3=\{q_3,q_5,q_7\}$ are three  $\{q_2\}$-home spaces and are not an $\{Init\}$-home space.
\\ 
$H_4=\{q_1, q_3,q_5\}$ is an $\{Init\}$-home space as well as a $\{q_2\}$-home space.
\\
Let us point out that there is no Init-home state.
}
\label{fig: inter-hs}
\end{figure}

It may be worth mentioning the straightforward following properties, given two subsets $A$ and $B$ of markings.
\begin{property}
\label{prop: inter-home}
    If $H$ is an A-home space, it is a B-home space for any nonempty subset $B$ of $A$. 
    If $H_1$ is an $A_1$-home space and $H_2$ is an $A_2$-home space, then $H_1 \cup H_2$ is an ($A_1 \cup A_2$)-home space.  
\end{property} 
However, the intersection of two home spaces is not necessarily a home space. 
Figure \ref{fig: inter-hs} represents a Petri net and its reachability graph with eight markings. $H_1, H_2$ and $H_3$, as defined Figure \ref{fig: inter-hs}, are three $\{q_2\}$-home spaces. 
While $H_1 \cap H_3 = \{q_3, q_5\}$ is a $\{q_2\}$-home space, $H_1 \cap H_2 = \{q_3\}$ is not a $\{q_2\}$-home space (even if it is a $\{q_3\}$-home state).
    
Given a Petri net $PN$ and a subset of markings $Init$, a \textit{sink} is a marking with no successor in the associated reachability graph $RG(PN,Init)$. 
More generally, a subset $S$ of markings is a sink in $RG(PN,Init)$ if and only if $RS(PN,S)=S$. Similarly, we say that strongly connected component $S$ of $RG(PN,Init)$ is \textit{strongly connected component sink} if and only if $\nexists\ y \in RS(PN,Init) \setminus S$ such that $\exists\ x \in S$ and $x \rightarrow y$.
As any directed graph, $RG(PN,Init)$ can have its vertices (markings) partitioned into strongly connected components which can be sink at the same time.
The following theorem can be found without proof in \cite{J87}
and allows determining whether a subset of $Q$ is an Init-home space considering that the set of strongly connected components sink is known (which is theoretically possible when $RG$ is finite). We believe that the corollary that follows is new. 

\begin{theorem}[Johnen]
\label{th: johnen}
$H$, a subset of $Q$  is an Init-home space if and only if $\forall S$ strongly connected component sink of $RG(PN,Init), S \cap H \neq \ $\O.
\end{theorem}
Let's $H$ be a subset of $Q$ such that for all strongly connected component sink $S$ of $RG(PN,Init),\ S \cap H \neq \ $\O. 
Let's consider a marking $q \in RS(PN,Init)$.
Then, there exists at least one strongly connected component sink $S$ of $RG(PN,q)$ such that $RS(PN,q) \cap S \neq $ \O. 
In other words, $\exists v \in RS(PN,q) \cap S $ such that $q \overset{*}{\rightarrow}v$. 
There also exists $h \in H \cap S$; $S$ being strongly connected, we have: $q \overset{*}{\rightarrow}v \overset{*}{\rightarrow}h$.
Hence, $H$ is an Init-home space.
The reverse is straightforward considering the definitions of home space, strongly connected component, and sink
\hfill
$\square$

Theorem \ref{th: johnen} allows us to directly derive the following three properties gathered in one corollary which completes the relationship between home spaces (or home states) and strongly connected component sink.
\begin{corollary}
\label{cor: sink-home-space}
If a marking $q_s$ of $RG(PN,Init)$ is a sink, then $q_s$ belongs to any init-home space.

If there exists a unique strongly connected component sink $S$ in $RG(PN,Init)$ then $S$ is a home space. 
Moreover, a marking is a home state if and only if it belongs to $S$.

If there exist two distinct strongly connected component sink, then there is no Init-home state.
\end{corollary}
The first statement of this corollary is similar to a property in \cite{JM91}. We deduce from the last statement of this corollary that the state machine $\langle PN, Init \rangle$ of Figure \ref{fig: inter-hs} has no Init-home state.

It is easy to prove that these properties hold even if the reachability graph is infinite, considering that the definitions of sources, sinks, or strongly connected components are the same as in the case where the directed reachability graph is finite. 
\begin{property}
\label{prop: home-state}
    we consider a Petri net $PN$ paired with a single initial marking $q_0$. Then, three following statements are equivalent: 
    \begin{itemize}
   \setlength\itemsep{0em}
    \item [(i)] the initial marking $q_0$ is a home state; 
    \item [(ii)] every reachable marking is a home state; 
    \item [(iii)] the reachability graph is strongly connected.
    \end{itemize}
\end{property} 
If $q_0$ is the initial marking, then, for all $ x, y \in RS(PN, q_0)$, there exist a path from $q_0$ to $x$ and a path from $q_0$ to $y$, and since $q_0$ is a home state, there also exists a path from $x$ to $q_0$ and from $y$ to $q_0$ in the reachability graph. 
Hence, $q_0, x$ and $ y$ belong to the same strongly connected component. 
We easily conclude that the reachability graph is strongly connected. 
The other elements of the property become obvious
\hfill
$\square$
The equivalence between the items (i) and (ii) can be found in \cite{J87,JM91}.
It is easy to deduce that any reachable marking from a home state $h$ is a home state even if $h$ is not the initial marking.
The strong connectivity of a given reachability graph means that some transitions are necessarily live.
 
This remark linking the notions of home state, strong connectivity of the reachability graph, and liveness requires further exploration. 
It is at the core of the following subsection.

\subsection{Home spaces, home states, and liveness}
\label{subsec: hs-liveness}
Home spaces together with invariants can greatly simplify the proof of fundamental properties of Petri nets (even including parameters as in \cite{BEISW20}) such as safeness, boundedness, or more complex behavioral properties such as liveness.  
Let us provide three properties supporting this idea. 

Let $\texttt{Dom(t)}$ denote the subset of markings from which the transition $t$ is executable, and $\texttt{Im(t)}$, the subset of markings that can be reached by the execution of $t$.
\begin{property}
\label{prop: liveness-domain-t}
A transition $t$ is live if and only if there exists a home space $H$ such that $H \subseteq \text{Dom(t)}$.

Moreover, if \texttt{Dom(t)} is a home space, then \texttt{Im(t)} is also a home space.

\end{property} 

This can be directly deduced from the usual definition of liveness and Definition \ref{def: homespace} of home spaces 
(if $H$ is a home space then $\forall q \in RS, \exists h \in H, q \overset{*}{\rightarrow}h$, 
and since $H \subseteq \text{Dom(t)}$, 
we have $q \overset{*}{\rightarrow}h \overset{t}{\rightarrow}$  )
\hfill
$\square$

We consider $\langle PN,Init \rangle$, a Petri net $PN$ with $Init$, its set of initial markings, its associated reachability set $RS$, its labeled reachability graph $LRG$, a home space $H$ such that $H \cap RS$ induces (see, for instance, \cite{Diestel10} for the notion of induced subgraph) a strongly connected subgraph of $LRG$.

The strong connectivity of the subgraph of the reachability graph induced by a home space allows us to reduce the condition of liveness for a transition from a universal one as in property \ref{prop: liveness-domain-t} to an existential one described as in the following lemma.
\begin{lemma}
\label{lem: live-t-home-space}
If a home space $H$ is such that $H \cap RS$ induces a strongly connected subgraph of LRG, then a transition $t$ is live if and only if there exist $ h_t \in H$ and $\sigma \in T^*$ such that $h_t \xrightarrow {\overset {\sigma t }{ }}$.
\end{lemma}
If $H$ is a home space, then $H \cap RS$ is also a home space, and for all $q \in RS$, there exist $ s_1 \in T^*$ and $h \in H$ such that $q \overset{s_1\ }{\rightarrow}h$.

The subgraph induced by $H \cap RS$ being strongly connected, there exists a path from $h$ to $h_t$; in other words, there exists $ s_2 \in T^*$ such that $h \overset{s_2\ }{\rightarrow} h_t$. 
We can construct a sequence $s=s_1s_2\sigma$ such that for all $ q \in RS, q\overset{st }{\rightarrow}$. 
Hence $t$ is live. 
The reverse is obvious
\hfill
$\square$



\begin{property}
\label{prop: home-state-liveness}
Let $PN$ be a Petri net and $q$ be a home state. 
Then, any transition that is enabled at $q$ is live in $\langle PN, Init \rangle$, and,
more generally,
a transition is live if and only if it appears as a label in $LRG(PN,q)$.
\end{property} 

This can be proven directly from the definition of liveness and Lemma \ref{lem: live-t-home-space}
\hfill 
$\square$

We can then deduce from this property that liveness is decidable for Petri nets equipped with a home state. More precisely, we have:

\begin{theorem}
\label{th: home-state}
Let $PN$ be a Petri net with a home state $q$, and $LCT(PN,q)$, the labeled coverability tree of $\langle PN, q \rangle$. 
A transition is live in $\langle PN, Init \rangle$ if and only if it appears as a label in $LCT(PN,q)$.
\end{theorem}

This can be proven directly from the fact that a transition appears as a label of an edge of $LRG(PN,q)$ if and only if it appears as a label of an edge of $LCT(PN,q)$, and by considering Property \ref{prop: home-state-liveness}
\hfill 
$\square$

\begin{corollary}
\label{cor: home-state}
For any Petri net $\langle PN, Init \rangle$, such that $Init$ is finite and  equipped with a home state, liveness is decidable.
\end{corollary}

This is a direct consequence of Theorem \ref{th: home-state} combined with Karp and Miller's theorem \cite{KM69}, stating that the coverability tree is finite, and the fact that $Init$ is finite
\hfill 
$\square$


\subsection{Semiflows, invariants, and home spaces}
\label{subsec: semiflows and home spaces}
Semiflows are intimately associated with home spaces.
We consider a logic formula $I_{f,q}$; its support $H(f,q)$ along definition \ref{def: inv}
We write: $H(f,q) = \{q \in Q\ \|\ q \models I \}$.


\begin{property}
\label{prop: inv-home-spaces-notflow}
Given a marking $q \in Q$ and $f \in \mathbb{Z}^d$,\\
the logic formula $I_{f,q}$ is an f-q-invariant if and only if its support $H(f,q)$ is a {q}-home space.\\
 Furthermore, $H(f,q) \cap H(g, q_)$ is a $\{q\}$-home space.
\end{property}
Equation (\ref{eq: inv}) is satisfied for all markings in $RS(PN,q)$ which is included in $H(f,q)$ considering $H(f,q)$  as a support or as a home space.
Note that $H(f,q)\ \cap \ H(g, q)$ is straightforwardly a $\{q\}$-home space, since they both contain $RS(PN,q)$\footnote{Let us recall that, in general, the intersection of home spaces is not a home space (see Figure \ref{fig: inter-hs}).} 
\hfill 
$\square$

Then, we can define $\Omega(q)$, the closure under $\cap$ of $\{ A \subseteq Q\ |\ \exists f \in \mathcal{F}^+, \ A = H(f,q)\}$, that is the smallest subset of $2^Q$ stable for $\cap$ containing $\{ A \subseteq Q\ |\ \exists f \in \mathcal{F}^+, \ A = H(f,q_0)\}$. 
For the same reason as for property \ref{prop: inter-home-spaces}, all elements of $\Omega$ are home spaces and there exists a unique nonempty minimal element $\omega(q)= \min_{A \in \Omega(q)} A$ that will be characterized in the next section.

From theorem \ref{th: inv-flow} and property \ref{prop: inv-home-spaces-notflow}, we easily deduce the following corollary:
\begin{corollary}
The three statements are equivalent:
\begin{itemize}
    \item $\forall q \in Q, I_{f,q}$ is an f-q-invariant,
    \item $f$ is a semiflow,
    \item $\forall q \in Q, H(f,q)$ is a {q}-home space.
\end{itemize}
\end{corollary}

Of course, it is possible to build a Petri net so that it becomes possible to find $f \in \mathbb{Z}^d, q \in Q$ such that $H(f,q)$ be a home space while $f$ is not a semiflow.
It is the case Figure \ref{fig: f-q-inv} where H(f,qi) is a {qi}-home space and we have seen that $f$ is not a semiflow.

\section{Generating sets and minimality}
\label{sec: generating sets}

Several results have been published, starting from the initial definition and structure of semiflows  \cite{M77} to a wide array of applications used especially to analyze Petri nets \cite{Colom2003,DworLo16,JACB18,Wol2019}.

Minimality of semiflows and minimality of their supports are critical to understand how to best decompose semiflows.
Invariants directly deduced from minimal semiflows relate to smaller weighted quantities of resources, simplifying the analysis of behavioral properties.
Furthermore, the smaller the support of semiflows, the more local their footprint (i.e., the more constrained the potential exchanges between resources is).
In the end, these two notions of minimality will foster analysis optimization.

\subsection{Three definitions}
\label{subsec: gs-basic-results}

\begin{definition}[Generating set]
\label{def: FG}
A subset $\mathcal{G}$ of $\mathcal{F^{+}}$ is a \textit{generating set over $\mathbb{S}$} if and only if for all $f \in \mathcal{F}^+$, 
we have $f = \sum_{g_i \in \mathcal{G}} \alpha_ig_i $, where $\alpha_i \in \mathbb{S}$ and $g_i \in \mathcal{G}$.
\end{definition}

Since $\mathbb{N} \subset \mathbb{Q^{+}} \subset \mathbb{Q}$, 
a generating set over $\mathbb{N}$ is also a generating set over $\mathbb{Q^{+}}$, and a generating set over $\mathbb{Q^{+}}$ is also a generating set over $\mathbb{Q}$. 
However, the reverse is not true and is, in our opinion, a source of some inaccuracies that can be found in the literature (see \cite{ColomTS2003}, for instance). 
Therefore, it is important to specify over which set of $\{\mathbb{N}, \mathbb{Q^{+}}, \mathbb{Q}\}$ the coordinates (used for the decomposition of a semiflow) vary.



Several definitions around the concept of minimal semiflow were introduced in \cite{STC98}, p. 319, in \cite{ColomTS2003},  p. 68, \cite{Kruck86}, \cite{CMPW09}, or in \cite{M78,M83}. However, we will only consider two basic notions in order theory: minimality of support with respect to set inclusion and minimality of semiflow with respect to the component-wise partial order on $\mathbb{N}{^d}$, since the various definitions found in the literature as well as the results of this paper can be described in terms of these two classic notions.

\begin{definition}[Minimal support]
A nonempty support $ \left \| f \right \| $ 
of a semiflow $f$ is \textit{minimal} with respect to set inclusion if and only if $\nexists \ g \in \mathcal{F}^+\setminus\{0\}$ such that $\left \| g \right \| \subset \left \| f \right \| $. 
\end{definition}

\begin{definition}[Minimal semiflow]
A non-null semiflow $f$ is \textit{minimal} with respect to $\leq$ if and only if $\nexists \ g \in \mathcal{F^{+}}\setminus\{0,f\}$ such that $g \leq f$.
\end{definition}

A minimal semiflow cannot be decomposed  as the sum of another semiflow and a non-null non-negative vector. 
This remark yields an initial insight into the foundational role of minimality in the decomposition of semiflows.
We are looking for characterizing generating sets such that they allow analyzing various behavioral properties as efficiently as possible.
That is to say that we want generating sets as small as possible and, at the same time, able to easily handle semiflows in $\mathcal{F}^+$.
First, the number of minimal semiflows over $\mathbb{N}$ can be quite large. 
Second, considering a basis over $\mathbb{Q}$ is of course relevant to handle $\mathcal{F}$, while less relevant when it is about $\mathcal{F}^+$, and may not capture behavioral constraints as easily.
We will have to consider $\mathbb{Q}^+$.

\subsection{Three decomposition theorems}
\label{subsec: 3theorems}

Generating sets can be characterized thanks to three decomposition theorems. 
A first version of them can be found in \cite{M78} with their proofs. 
A second version can be found in \cite{M23} with improvements.  
Here, Theorem \ref{th: over N}, which is valid over $\mathbb{N}$,
is extended to a necessary and sufficient condition that characterizes a minimal semiflow and generating sets over $\mathbb{N}$. 
This result is provided with a new proof using Gordan's lemma (see Lemma \ref{lem: Gordan}).
Theorems \ref{th: min support} and \ref{th: decomp} are recalled for completeness and are unchanged from \cite{M23} where additional results can be found about minimality.

\subsubsection{Decomposition over non-negative integers}

The fact that there exists a finite generating set over $\mathbb{N}$ is non-trivial and is often taken for granted in the literature on semiflows. 
In fact, this result was proven by Gordan, circa 1885, then Dickson, circa 1913. 
Here, we directly rewrite Gordan's lemma \cite{AB86} by adapting it to our notations.

\begin{lemma} (\textbf{Gordan})
\label{lem: Gordan}
Let $\mathcal{F^+}$ be the set of non-negative integer solutions of the System of equations (\ref{eq: inv-semiflow}). Then, there exists a finite generating set over $\mathbb{N}$ of semiflows in $\mathcal{F^+}$.
\end{lemma}

The question of the existence of a finite generating set being solved for $\mathbb{N}$, is necessarily solved for $\mathbb{Q^+}$ and $\mathbb{Q}$. 
Lemma \ref{lem: Gordan} is necessary not only to prove the decomposition theorem but also to claim the computability of the extremums described in Theorem \ref{th: bounds}.

\begin{theorem} (\textbf{Decomposition over $\mathbb{N}$})
\label{th: over N}
A semiflow is minimal if and only if it belongs to all generating sets over $\mathbb{N}$.

The set of minimal semiflows of $\mathcal{F^{+}}$ is a finite generating set over $\mathbb{N}$.
\end{theorem}
Let's consider a semiflow $f \in \mathcal{F^{+}}\setminus\{0\}$ and its decomposition over any family of $k$ non-null semiflows $f_i, 1 \leq i \leq k$. Then, there exist $ a_1,...,a_k \in \mathbb{N}$ 
such that $f = \sum_{i=1}^{i=k}a_if_i$. 
Since $f \neq 0$ and all coefficients $a_i$ are in $\mathbb{N}$, there exists $ j \leq k$ such that $0 \lneq f_j \leq a_jf_j \leq f$. If $f$ is minimal, then $a_j=1$ and $f_j=f$.
Hence, if a semiflow is minimal, then it belongs to any generating set over $\mathbb{N}$. The reverse will become clear once the second statement of the theorem is proven.

Applying Gordan's lemma, there exists a finite generating set, $\mathcal{G}$. 
Since any minimal semiflow is in $\mathcal{G}$, the subset of all minimal semiflows is included in $\mathcal{G}$ and therefore finite. Let $\mathcal{E} = \{e_1,...e_n\}$ be this subset and prove by construction that $\mathcal{E}$ is a generating set. 

For any semiflow $f \in \mathcal{F^+}$, we 
build the following sequence leading to the decomposition of $f$ over $\mathcal{E}$:

i) $r_0 = f$,

ii)  $r_i = r_{i-1} - k_ie_i$ such that $ r_i \in  \mathcal{F^+}$
and $r_{i-1} - (k_i+1)e_i \notin \mathcal{F^+}.$

By construction of the non-negative integers $k_i$, we have $r_n = f- \sum_{i=1}^{i=n} k_ie_i \in \mathcal{F^+}$
and $\forall e_i \in \mathcal{E},\ r_n-e_i \notin  \mathcal{F^+}$
therefore, $\forall e_i \in \mathcal{E},\ \exists j, (r_n)_j - (e_i)_j < 0$;
therefore  $\nexists  e_i \in \mathcal{E}$ such that $e_i \leq r_n$. 
This means that $r_n$ is either minimal or null. 
Since $\mathcal{E}$ includes all minimal semiflows, therefore 
$r_n=0$, and any semiflow can be decomposed as a linear combinations of minimal semiflows; in other words, $\mathcal{E}$ is a finite generating set\footnote{If $\mathcal{E}$ were to be  infinite, the construction could still be used, since the monotonically decreasing sequence $r_i$ is bounded by 0 and $\mathbb{N}$ is nowhere dense, so we would have
$ \lim \limits_{n \to \infty} (f- \sum_{j=1}^{j=n}k_je_j) = 0$, with the same definition of the coefficients $k_j$ as in ii).}.

It is now clear that if a semiflow $f$ belongs to any generating set, then it belongs in particular to $\mathcal{E}$; therefore, $f$ is a minimal semiflow
\hfill
$\square$

Let's point out that since $\mathcal{E}$ is not necessarily a basis, the decomposition is not unique in general and depends on the order in which the minimal semiflows of $\mathcal{E}$ are considered to perform the decomposition.

However, a minimal semiflow does not necessarily belong to a generating set over $\mathbb{Q^{+}}$ or $\mathbb{Q}$. 
\subsubsection{Decomposition over semiflows of minimal support}
These two theorems can already be found in \cite{M23}.
\begin{theorem} (\textbf{Minimal support})
\label{th: min support}
If $I$ is a minimal support, then 

i) there exists a unique minimal semiflow $f$ such that $I = \left \| f \right \|$ and, for all $ g \in \mathcal{F^{+}}$ such that $\left \| g \right \| = I$, there exists $ k \in \mathbb{N}$ such that $g = kf$, and

ii) any non-null semiflow $g$ such that $\left \| g \right \| = I$ constitutes a generating set over $\mathbb{Q^{+}}$ or $\mathbb{Q}$ for $\mathcal{F}_I^+ = \left\{ g \in \mathcal{F^+}\:|\ \left\| g\right\|= I\right\}$.
\end{theorem}
In other words, $\{f\}$ is a unique generating set over $\mathbb{N}$ for $\mathcal{F}_I^+ =\{g \in \mathcal{F^{+}}\;|\left \| g \right \| = I\}$. 
Indeed, this uniqueness property is lost in $\mathbb{Q^{+}}$ or in $\mathbb{Q}$, since any element of $\mathcal{F}_I^+$ is a generating set of $\mathcal{F}_I^+$ over $\mathbb{Q^{+}}$ or $\mathbb{Q}$.

\begin{theorem} (\textbf{Decomposition over $\mathbb{Q}^+$})
\label{th: decomp}
Any support $I$ of semiflows is covered by the finite subset $\{I_1, I_2, \dots, I_N\}$ of minimal supports of semiflows included in $I$:
$I = \bigcup_{i=1}^{i=N} I_i$.

Moreover,
for all $ f \in \mathcal{F^{+}}$ such that $\left \| f \right \| \subseteq  I$, one has $f=\sum_{i=1}^{i=N} \alpha_ig_i$, where, for all $  i \in \{1,2,...N\},\ \alpha_i \in \mathbb{Q^{+}}$ and the semiflows $g_i$ are such that $\left \| g_i \right \| = I_i$.
\end{theorem}

A sketch of the proof of Theorem \ref{th: decomp} using Property \ref{prop: support-plus-union} can be found in \cite{MR79}, and a complete proof, in \cite{M78}.
This last theorem says that one cannot have a generating set with less than $n$ semiflows where $n$ is the number of minimal supports included in $P$.

\subsection{Three extremums computability}
\label{subsec: bounds}

Knowledge of any finite generating set allows a practical computation of the three extremums ($\lambda$, $\theta$, and $\omega$) defined in the previous sections. 



We know that a finite generating set does exist by Gordan's lemma (\ref{lem: Gordan}) and we know how to compute a generating set (see \cite{martinez1982} for instance). Subsequently, we can state the following theorem which expresses the fact that three extremums ($\lambda$, $\theta$, and $\omega$) are computable as soon as any finite generating set is available.  
\begin{theorem}
\label{th: bounds}
Let $\mathcal{E} = \{e_1,...e_N\}$ be any finite generating set of $\mathcal{F}^+$, and $q_0 \in Q$, an initial marking.
\begin{itemize}
  \item[(i)]  If $\mathcal{E}$ is over $\mathbb{S}$, then we have:
  $\omega(q_0) = \bigcap_{f \in \mathcal{F^+}} H(f,q_0) = \bigcap_{e_i \in \mathcal{E}} H(e_i,q_0)$;
  \item[(ii)] If $\mathcal{E}$ is over $\mathbb{Q}^+$ or $\mathbb{N}$, then, for any place $p$ belonging to at least one support of a semiflow of $\mathcal{F}^+$, for all $q \in RS(PN,q_0)$, we have :   
$$q(p) \leq \lambda(p,q_0) = \min_{\{f\in \mathcal{F}^+\ |\ f(p) \neq 0\}} \frac{f^\top q_0}{f(p)} = \min_{\{e_i\in \mathcal{E}\ |\ e_i(p) \neq 0\}} \frac{{e_i}^\top q_0}{e_i(p)}; $$

  \item[(iii)] If $\mathcal{E}$ is over $\mathbb{Q}^+$ or $\mathbb{N}$, then, for any transition $t$ belonging to at least one support of a semiflow of $\mathcal{F}^+$, for all $q \in RS(PN,q_0)$, we have :   
$$\theta(t,q_0) = \min_{\{f\in \mathcal{F}^+\ |\ f^\top Pre(\cdot,t) \neq 0\}\}}
\frac{f^\top q_0 }{f^\top Pre(\cdot,t)} = \min_{\{e_i\in \mathcal{E}\ |\ e_i^\top Pre(\cdot,t) \neq 0\}\}}
\frac{f^\top q_0 }{f^\top Pre(\cdot,t)}$$.
 
\end{itemize}
\end{theorem}


Regarding item (i), let's consider 
$f \in \mathcal{F^+}$ with $f = \sum_{i=1}^{i=N} \alpha_ie_i$ and $q \in \bigcap_{e_i \in \mathcal{E}} H(e_i,q_0)$. Then, 
$\alpha_i(e_i^\top q) = \alpha_i(e_i^\top q_0)$, for all  $i\in \{1,...N\}$, and, hence
$ \sum_{i = 1}^{i = N} \alpha_i(e_i^\top q) = \sum_{i = 1}^{i = N} \alpha_i(e_i^\top q_0)$.
Then, for all $f \in \mathcal{F^+}$, $f^\top q = f^\top q_0$, 
and $q \in H(f,q_0)$.

Therefore, since $\mathcal{E} \subset \mathcal{F^+}$ directly implies $(\bigcap_{f \in \mathcal{F^+}} H(f,q_0))  \subseteq \bigcap_{e_i \in \mathcal{E}} H(e_i,q_0)$, we have:
$\bigcap_{e_i \in \mathcal{E}} H(e_i,q_0) = \bigcap_{f \in \mathcal{F^+}} H(f,q_0) = \omega(q_0)$.

For the item (ii), let's consider a marking $q_0$, a place $p$, and a semiflow $f$ of $\mathcal{F}^+$ such that $f(p) > 0$ and $f = \sum_{i=1}^{i=N}\alpha_i e_i$, where $\alpha_i \geq 0$, for all $i \in \{1,...N\}$.

Let's 
define $\lambda_\mathcal{E}$ such that
$\lambda_\mathcal{E} = 
\min_{\{e_i\in \mathcal{E}\ |\ e_i(p) \neq 0\}} \frac{{e_i}^\top q_0}{e_i(p)}$.
Then, there exists $j$ such that $1 \leq j \leq N$ and $\lambda_\mathcal{E} = \frac{{e_j}^\top q_0}{e_j(p)}$.

Therefore, for all $ i \leq N$ such that $e_i(p) \neq 0$,  
there exists $\delta_i \in \mathbb{Q}^+$ such that:
$\frac{{e_j}^\top q_0}{e_j(p)} = \frac{{e_i}^\top q_0 - \delta_i}{e_i(p)}$. It can then be deduced, for all $i$ such that $\alpha_ie_i(p) \neq 0$:
$$\lambda_\mathcal{E} =  \frac{\alpha_i({e_i}^\top q_0 - \delta_i)}{\alpha_i e_i(p)},$$
and, therefore:
$$
\lambda_\mathcal{E} = \frac{\sum_{\{i\ |\ \alpha_ie_i(p) > 0\}} \alpha_i({e_i}^\top q_0 - \delta_i)}{\sum_{\{i\ |\ \alpha_ie_i(p) > 0\}}\alpha_i e_i(p)} 
 $$
Since $\sum_{\{i\ |\ \alpha_ie_i(p) > 0\}} \alpha_i({e_i}^\top q_0) = f^\top q_0 - \sum_{\{i\ |\ \alpha_ie_i(p) = 0\}} \alpha_i({e_i}^\top q_0)$ 

and $\sum_{\{i\ |\ \alpha_ie_i(p) > 0\}}\alpha_i e_i(p) = f(p)$

$$
\lambda_\mathcal{E} 
= \frac{f^\top q_0 -\sum_{\{i | e_i(p)>0\}}\alpha_i\delta_i - \sum_{\{i\ |\  e_i(p) = 0\}}\alpha_i{e_i}^\top q_0}{f(p)} 
$$

Then, since $\delta_i \geq 0$ and $\alpha_i \geq 0$ for all $i$ such that $1\leq i \leq N$,
$$\lambda_\mathcal{E} \leq \frac{f^\top q_0 }{f(p)}$$
This being verified for any semiflow of $\mathcal{F}^+$, we have $\lambda(p,q_0) = \lambda_\mathcal{E}$.

The item (iii) of the theorem can be proven by a similar demonstration
\hfill
$\square$

Item (i) is similar to a result that can be found in \cite{J87}.
The complexity of computing item (ii) or (iii) depends on $N$ the number of elements of $\mathcal{E}$. We know from theorem \ref{th: decomp} that $N$ cannot be less than the number of minimal supports.


\section{Reasoning with invariants, semiflows, and home spaces}
\label{sec: ex}

Invariants, semiflows, and home spaces can be used in combination to ease the proof of a rich array of behavioral properties of Petri nets, 
in particular when using parameters. 

\subsection{Methodically analyzing behavioral properties}
\label{subsec: method}
The results presented in this paper provide verification engineers with a few steps to methodically analyze and prove behavioral properties, in particular, that a subset of transitions are live:
\begin{itemize}
    \item [\textbf{-}] reduce the model using reduction rules such as in\cite{Berth86,STC98}, paying attention at preserving the support of properties that we seek to prove,
    \item [\textbf{-}] run existing algorithms to compute Generating sets,\cite{martinez1982,AM82}, even with parameters as in \cite{HaddadCouvreurPP91}; (comparisons and benchmarks can be found in \cite{ColomS89} and more recently in \cite{YTM23}),
    \item [\textbf{-}] from a generating set, using property \ref{prop: inter-home-spaces}, infer a first home space that concisely describes how tokens are distributed over places,
    \item [\textbf{-}] use theorem \ref{th: bounds} to prune away impossible situations,
    \item [\textbf{-}] step by step proceed by refinement of home states, by finding which transitions or which arithmetic transformation can be enabled and constrain current home spaces until reaching a possible home state,  
    \item [\textbf{-}] Use algorithms as in \cite{KM69,FinkelHK20} to construct the labeled coverability tree (LCT), and then deduce which transitions are live from the ones that appears in LCT (theorem \ref{th: home-state}),
    \item [\textbf{-}] ultimately, decide whether the Petri net is live or not.
\end{itemize}



Here, through two related parameterized examples, we proceed by using basic arithmetic and some particularity of the structure of the model to determine a home space and a home state in the second case. Then, it becomes easy to determine for which values of the parameters the Petri net possesses the required liveness property.

First, we propose to look at an example with a parameter $i$ to define its $\mathit{Pre}$ and $\mathit{Post}$ functions. This example allows one to detect whether a natural number $n$ is a multiple of $i$. 
The second example is an extension of the first one with a coloration of the tokens allowing one to detect the remainder of the Euclidean division of $n$ by $i$.  
\subsection{A first example}
\label{subsec: first-exemple}
\begin{figure}[ht]
\centering
\includegraphics[width=0.5\textwidth]{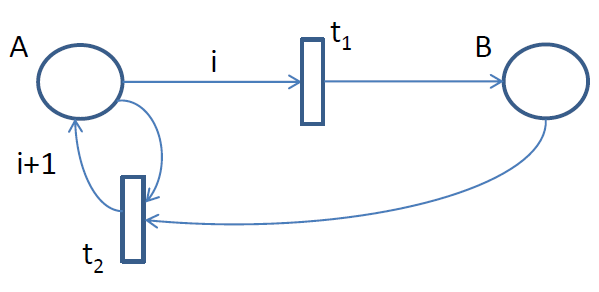}
\caption{
Semiflows must verify the system of equations (\ref{eq: inv-semiflow}) which is reduced to the following equation: $i \times a = b$, for which $g^\top  = (1,i)$ is an obvious solution.
$TN(i)$ is live if and only if $g^\top q_0 > i$ and is not a multiple of $i$, regardless  of the initial marking of $B$. 
For $i=1$, $TN(1)$ has no live transition, regardless of the initial marking.}
\label{fig: tinyi}
\end{figure}

The Petri net $TN(i) = \langle \{A,B\}, \{t_1,t_2\}, Pre, Post \rangle$ in Figure~\ref{fig: tinyi} is defined by: 

$Pre(\cdot,t_1)^\top =(i,0); Pre(\cdot,t_2)^\top =(1,1)$;

$Post(\cdot,t_1)^\top =(0,1); Post(\cdot,t_2)^\top =(i+1,0).$

The initial marking $q_0$ is such that $q_0(A)=n$ and $q_0(B)=x$, where $n$ and $x \in  \mathbb{N}$.

A first version of this example can be found for $i=2$ in \cite{BR82} or in \cite{M83}, without proof. 
Here, the Petri net $TN(i)$ is enriched by introducing a parameter $i$ such that $i>1$ in order to define its Pre and Post functions.

$g^\top  = (1, i)$ is the minimal semiflow of minimal support, and we can prove that $\left\langle TN(i),q_0 \right\rangle$ is not live if and only if $g^\top q_0 \leq i$ or $g^\top q_0 = n \times i$, independently of $q_0(B)$.
In other words, $TN(i)$ recognizes whether a given number $n$ is a multiple of $i$.

First, if $g^\top q_0 < i$, then the enabling threshold of $t_1$ can never be reached (Property \ref{prop: enabling-threshold}) and neither $t_1$ nor $t_2$ can be executed (since $q_0(B)$ is necessarily null to satisfy the inequality). 
Second, if $g^\top q_0 \geq i$, then we consider the Euclidean division of $g^\top q_0$ by $i$, giving $g^\top q_0 = n \times i + r$, where $r < i$. Then, since $g$ is a semiflow, $g^\top q = q(A)+iq(B) \equiv r \mod i$, and therefore, $q(A) \equiv r \mod i$, for all $q \in RS(TN(i),q_0)$. If $r=0$,  then we have $q(A)= n \times i - i \times q(B)$, and $t_1$ can always be executed $n - q(B)$ times to reach a marking with zero token in $A$. 

If $r \neq 0$ and $g^\top q_0 > i$, then after executing $t_1$ $n - q(B)$ times, we reach a marking with $r$ tokens in $A$ and at least one token in $B$. 
Therefore,
$ H =\{q \in RS(TN(i),q_0) \ |\ q(A) \neq 0\ \land\ q(B) \neq 0\}$ is a home space such that $H \subseteq \texttt{Dom($t_2$)}$ so we can apply property \ref{prop: liveness-domain-t} proving that $t_2$ is live.
It is easy to conclude that the Petri net $TN(i)$ is live if and only if $g^\top q_0 > i$ and is not a multiple of $i$, regardless of the initial marking of $B$
\hfill 
$\square$ 

We can point out that it was not necessary to develop a symbolic reachability graph in order to decide whether or not the Petri net is live or bounded. 
We could analyze the Petri net even partially ignoring the initial marking (i.e., the analysis could be conducted independently of the values taken by $q_0(B)$).

\subsection{Euclidean division}
\label{subsec: euclidean}
From the properties of $TN(i)$, it is natural to progress by one more step and propose to design a Petri net with the ability not only to recognize whether a natural number $n$ is a multiple of a given natural number $i$, but more generally to recognize the remainder of the Euclidean division of $n$ such that $n>0$ by $i$ such that $i>1$.

\begin{figure}[ht]
\centering
\includegraphics[width=0.47\textwidth]{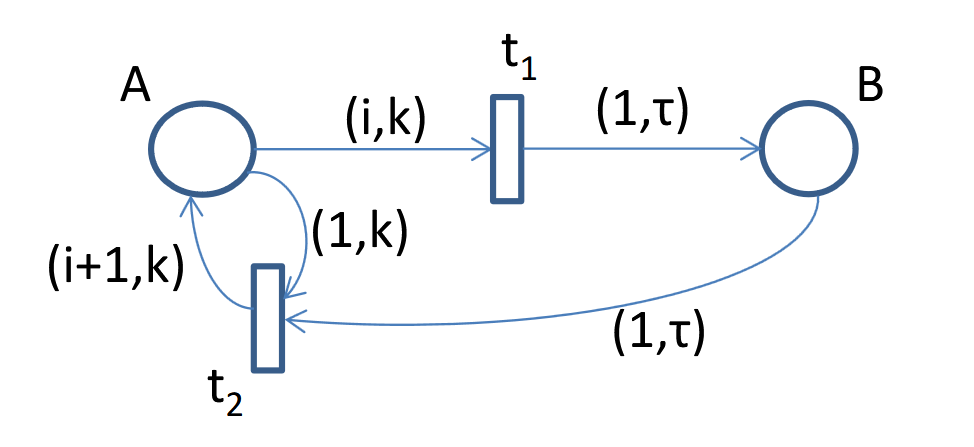}
\includegraphics[width=0.47\textwidth]{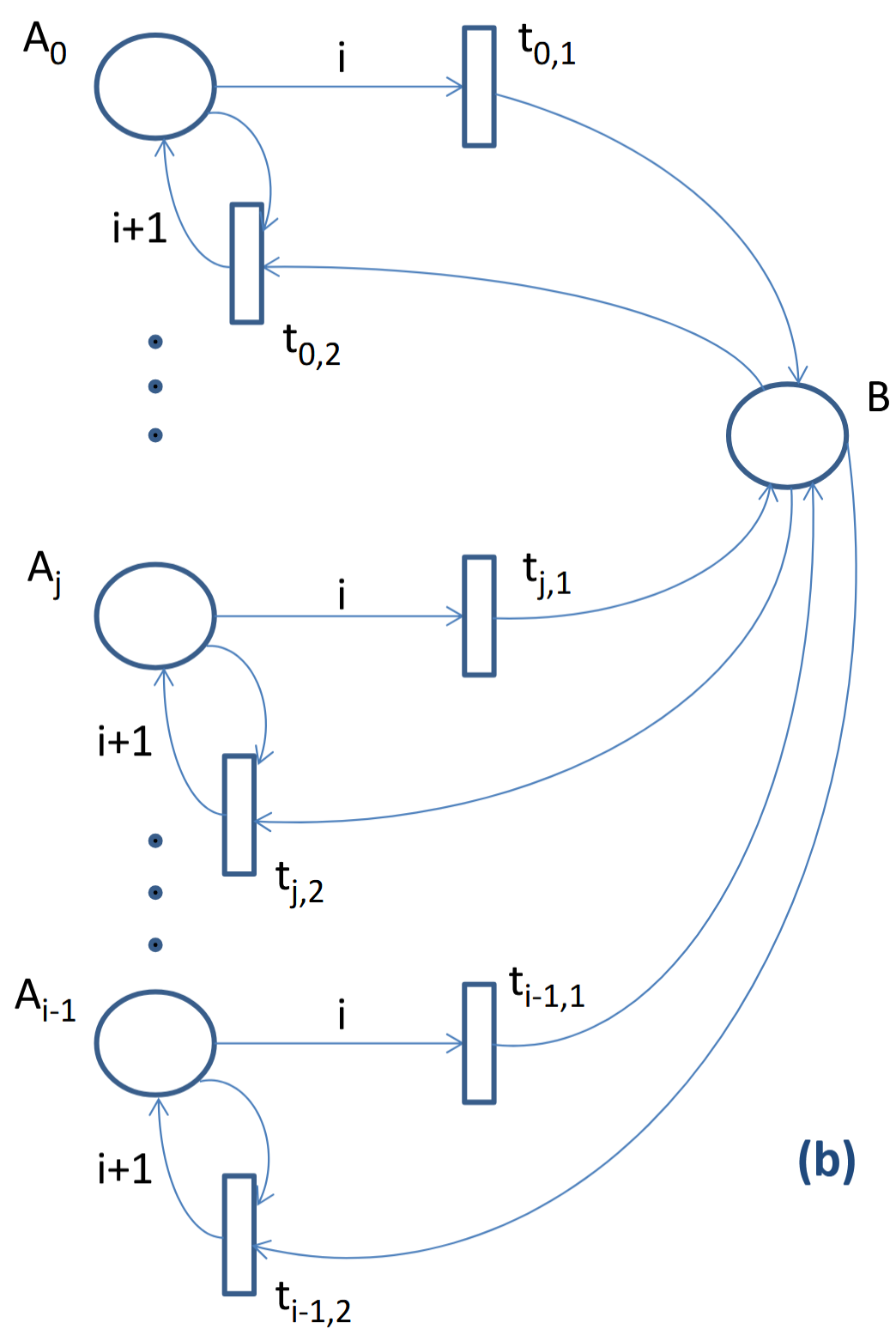}
\caption{$TNCED(i)$ of Figure \textbf{(a)}, is a Colored Petri net with a set $C$ of colors to distinguish values between 0 and $i-1$ plus $\tau$ as an undefined color of token; 
$\ C = ([0,...i-1]\cap \mathbb{N})\cup \{\tau\} $. 
\\
We have a system of $i$ equations: 
$i \times a_j = b$ with $j \in \left[0,i-1\right]\cap \mathbb{N}$, 
for which $g$ such that $g(A(j)) = 1$ for $j \in \left[0,i-1\right]\cap \mathbb{N}$ and $g(B(\tau))=i$ is the same minimal semiflow of minimal support in $\mathbb{N}$.
\\
$TNED(i)$ of Figure \textbf{(b)} is the unfolded version of $TNCED(i)$.
To compute semiflows, we have the same system of $i$ equations: 
$i \times a_j = b$ with $j \in \left[0,i-1\right]$, 
for which $g$ such that $g(A_j) = 1$ for $j \in \left[0,i-1\right]$ and $g(B)=i$ is as expected, the same minimal semiflow of minimal support in $\mathbb{N}$ as for $TNCED(i)$.
\\
This parameterized Petri net (colored or unfolded) allows knowing the remainder of the Euclidean division of a natural number $n$ by $i$.
}
\label{fig: euclidean}
\end{figure}

To this effect, we first consider the Colored Petri net $TNCED(i)$ of Figure \ref{fig: euclidean}(a), and the parameter $i \geq 2$. 
Second, to simplify the reasoning, we unfold $TNCED(i)$ into the classic Petri net $TNED(i)$, where each place $A_j$ represents the place $A$ with the color $j$ of $TNCED(i)$ (as shown Figures \ref{fig: euclidean}(a) and \ref{fig: euclidean}(b)). 

$TNED(i)= \langle P,T, Pre, Post \rangle$ is such that:

$P = \{\{A_j |\ j \in \left[0 , i-1\right]\}, B\}$ and  
$T=\{t_{j,1} ,t_{j,2} |\ j \in \left[0 , i-1\right]\}$, where $\mathit{Pre}$ and $\mathit{Post}$ are defined for $j \in  \left[0 , i-1\right]\ $by :

\noindent $Pre(A_j,t_{j,1})=i,\ Pre(B,t_{j,1})=\ Pre(A_j,t_{j,2})=1, Pre(p,t)=0$ otherwise,

\noindent $Post(A_j,t_{j,2})=i+1, Post(B,t_{j,1})=1, Post(p,t)=0,$ otherwise. 


The initial marking is such that $q_0(A_j) = n+j$, where $j \in \left[0 , i-1\right]$, and $q_0(B) = x$, where $n>0$ and $x$ are natural numbers.


$g^T = (1, \cdots 1,i)$, such that $g(A_j) = 1$ for $j \in \left[0,i-1\right]$, and $g(B)=i$ is the sole minimal semiflow of minimal support in $\mathbb{N}$.
We have a first associated invariant $I_{g,q_{0}}$, since:
$$\forall q \in RS(TNED(i),q_0):
  g^{\top}q_0 = g^{\top}q
= \sum_{j=0}^{j=i-1}q_0(A_j) + i q_0(B)
= i \times (x + n + \frac{i-1}{2}).$$

Then, we need to notice that any place $A_j$ is connected to exactly two transitions, $t_{j,1}$ and $t_{j,2}$, such that:

$Post(A_j,t_{j,1})-Pre(A_j,t_{j,1}) = -i$,

$Post(A_j,t_{j,2})-Pre(A_j,t_{j,2}) = i$.

Hence, for all $ q \in RS(TNED(i),q_0)$ and $j \in \left[0,i-1\right]$, $q(A_j)$ can only vary by $\pm i$. 
We then deduce a family of $i$ invariants $I(j)$ for $j \in \left[0,i-1\right]$: 
$$I(j): \forall q \in RS(TNED(i),q_0),
\  q(A_j) \equiv q_0(A_j) \mod i.$$
By performing the Euclidean division of $q_0(A_j)$ by $i$, we have: $$q_0(A_j) = n + j = a_j \times i + \alpha_j, where\ \alpha_j < i,\ \forall
 j \in \left[0,i-1\right]$$ 
A new family of $i$ invariants $I'(j)$ can be directly deduced from each $I(j)$ and the remainder of each Euclidean division:

$$I'(j): \forall q \in RS(TNED(i),q_0),\  q(A_j) \geq \alpha_j.$$

Furthermore, it must be pointed out that $\{\alpha_0, \cdots \alpha_{i-1}\}$ is a permutation of $\{0, \cdots i-1\}$. 
Indeed, if there exist $ j < i$ and $ j'<i$ such that $\alpha_j = \alpha_{j'}$, then $n+j - a_j \times i = n+j' - a_{j'}\times i $, and $|j-j'|=|a_j-a_{j'}|\times i$. Since $|j-j'|<i$, we have $a_j=a_{j'}$ and $j=j'$. Therefore:
\begin{itemize}
    \item[(a)] $\forall q \in RS(TNED(i),q_0),\ \sum_{j=0}^{j=i-1}q_(A_j) \geq \frac{i(i-1)}{2}$ (directly from the $I'(j)$ family of invariants) and
    \item[(b)] there is a unique $k \in \left[0,i-1\right]$ such that $\alpha_k = 0$.
\end{itemize}
From (a) and $I_{g,q_{0}}$, we deduce that $\forall q \in RS(TNED(i),q_0),\ q(B) \leq x+n$ (which by the way, is a better bound than the one that can be deduced from Proposition \ref{prop: boundedness-alg}). Also, from $I(j)$, we can deduce, $\forall q \in RS(TNED(i),q_0), \ q(A_j)=y_j \times i + \alpha_j,$
where $y_j \in  \mathbb{N}$ and $\alpha_j$ is the same remainder as for the Euclidean division of $q_0(A_j)$.

From any reachable marking $q$, the sequence $\sigma_q = t_{0,1}^{y_{0}} \cdots t_{i-1,1}^{y_{i-1}}$ can be executed and reach the marking $q_h$ such that, for all $ j \in \left[0,i-1\right],\ q_h(A_j)=\alpha_j$ and $q_h(B)=x+n$.

$q_h$ is a home state since $\sigma_q$ is defined for any reachable marking (Note that $q_0$ is not a home state since $q_0 \neq q_h$ (since $n > 0$) and that $q_0$ is unreachable from $q_h$).

From Property \ref{prop: home-state-liveness}, we deduce that, since any transition $t_{j,2}$ where $j \neq k$ is executable ($n>0$ hence, $q_h(B) > 1$ and $q_h(A_j) = \alpha_j > 0$), then $t_{j,2}$ is live, and, therefore, the corresponding transitions $t_{j,1}$ are also live.

From (b), $q_h(A_k) = 0$ from, which we deduce that $t_{k,1}$ and $t_{k,2}$ are not live\footnote{Actually, it suffices to notice that $A_k$ is an empty siphon (as defined section \ref{subsec: Petri nets definitions}) that once empty remains empty.}.
Finally, we have $n+k = a_k \times i$, and the remainder of the Euclidean division of $n$ by $i$ is $i-k$.

$TNED(i)$ provides the ability to recognize this remainder by the remarkable fact that $(t_{k,1},t_{k,2})$ is the only couple of transitions not live
\hfill 
$\square$ 

\begin{figure}[ht]
\centering
\includegraphics[width=0.5\textwidth]{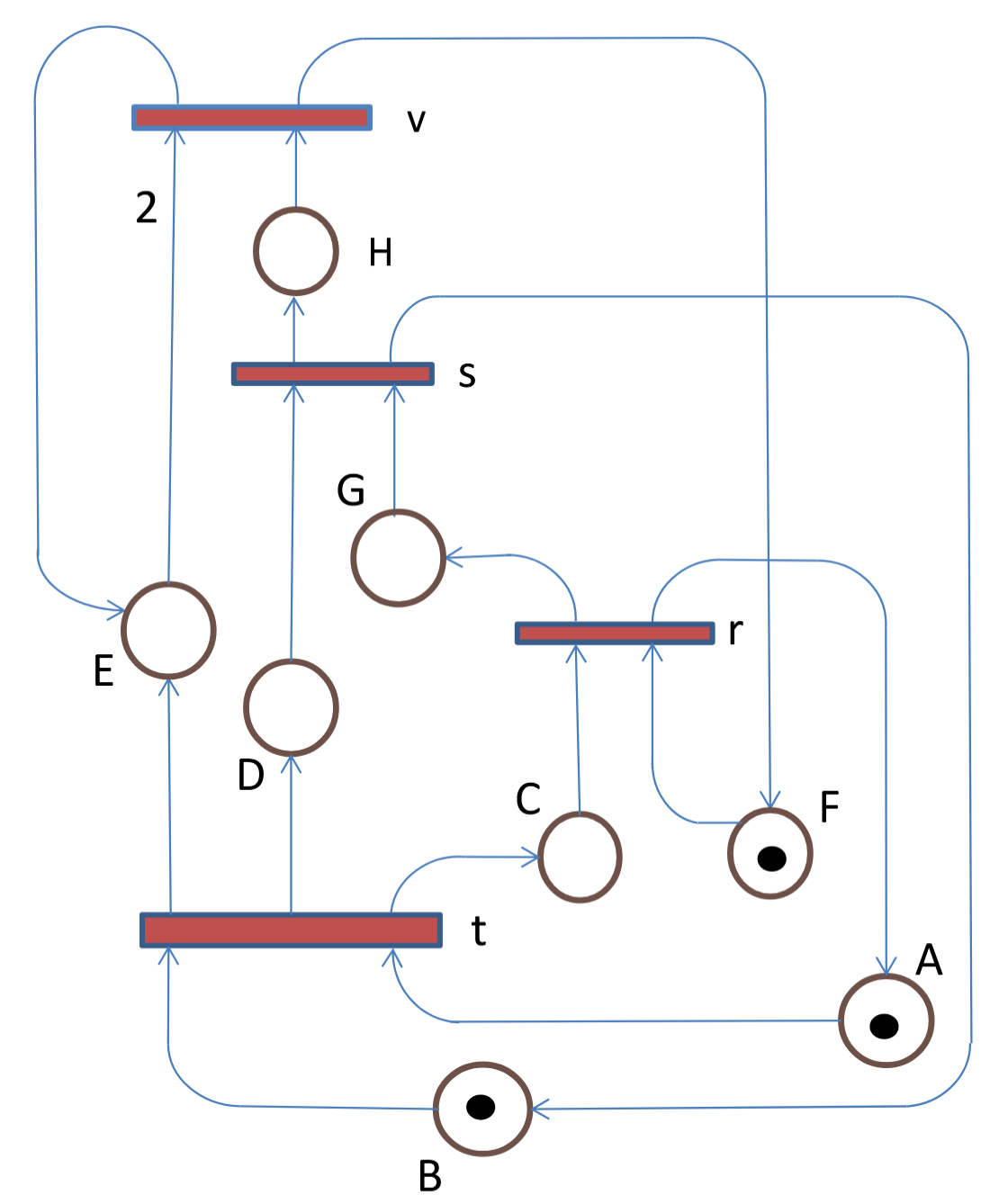}
\caption{
The initial state $q_0$ is such that: $q_0(A)=q_0(B)=q_0'F)=1, q_0(p)=0$ otherwise.
\\
This Petri net is live and for instance, $q_c$ such that: $q_c(C)=q_c(D)=q_(E)=q_c(F)=1, q_c(p)=0$ otherwise, is a home state.}
\label{fig: Home-state-c}
\end{figure}

Most of the time, in real-life use cases, when a model accepts a set of home states, the initial marking belongs to it. 
This is not the case in our example, where the initial marking $q_0$ is not a home state, and $q_h$ is a home state.

This could suggest (as conjectured in \cite{Me25}) that 
if the initial marking $q_0$ of a Petri net $PN$ is not a home state and there exists a home state $q_h$ in $RS(PN,q_0)$, 
then there exists at least one non-live transition in a sequence of transitions from $q_0$ to $q_h$.
This statement holds for state machines, however, it does not hold in the general case, as shown by the Petri net Figure \ref{fig: Home-state-c}.

\section{Conclusion}
\label{sec: concl}

It has been recalled how semiflows allow inferring strong constraints over not only all reachable markings but all possible markings of a given Petri net, which greatly help analysis of behavioral properties such as not only boundedness but also liveness.
This allows pursuing the same analysis on a set of initial markings (defined by a parameter).
Moreover, analysis can be performed with incomplete information,
particularly when markings and even structures are described with parameters as in our two examples.

The set of semiflows can be characterized with the notion of minimal generating set, and we hope that our three decomposition theorems reached their final version.
They were useful to make properties on boundedness or liveness computable.
Theorem \ref{th: bounds} consolidates our thesis saying that the information deduced from a generating set brings most of if not all the information that semiflows in $\mathcal{F}^+$ can provide for analyzing behavioral properties.

Most of the time, especially with real-life system models, it will be possible to avoid a painstaking symbolic model checking or a parameterized and complex development of a reachability graph \cite{DRvB01,ChiolaDFH97}. 

We introduced new results about home spaces; in particular, theorem \ref{th: home-state} is new to the best of our knowledge (for instance, it does not appear in the recent survey on decidability issues for Petri nets \cite{EN24}). This theorem is interesting for at least two different reasons. 
First, from a theoretical point of view, it characterizes a class containing unbounded Petri nets (the existence of a home state does not involve that the Petri net is bounded). Second, from a practical point of view, it shed a new light on the usage of coverability graphs, since real-life systems often have a home state by design. 
It increases the importance one can grant to the construction and the optimization of coverability trees, which is used mostly to determine which places are bounded (see important works by Finkel and al \cite{FinkelHK20} about accelerating this construction) by supporting the analysis of liveness.

At last, we presented most of these results in the framework of Petri nets, we believe that part of these results apply to Transition Systems, especially, section \ref{sec: HS} on home spaces . 
This, indeed, constitutes a starting point for future work. 

\bibliography{bibfile}

\end{document}